\documentclass[11pt,a4paper]{article}

\usepackage{graphics}
\usepackage{graphicx,epsfig}
\usepackage{epsfig}
\usepackage{graphicx}
\usepackage{a4wide}
\usepackage{amsmath}
\usepackage{amsfonts}
\usepackage{enumerate}
\begin{document}

\newcommand{\pf}{{\bf Proof : }}
\newtheorem{definition}{Definition}
\newtheorem{theorem}{Theorem}[section]
\newtheorem{lemma}{Lemma}[section]
\newtheorem{proposition}{Proposition}[section]
\newtheorem{remark}{Remark}[section]
\newtheorem{corollary}{Corollary}[section]
\newtheorem{example}{Example}[section]

\newcommand{\ind}{1\hspace{-2.3mm}{1}}
\newcommand{\qed}{\hfill{\ \ \rule{2mm}{2mm}} \vspace{0.2in}}

\title{\bf
Classification and Regression Error Bounds for Inhomogenous Data With Applications to Wireless Networks
}
\author{Ghurumuruhan Ganesan\\
IISER Bhopal,\\
e-mail : gganesan82@gmail.com\\
}
\date{}

\maketitle

\thispagestyle{empty}

\begin{abstract}
\noindent 

In this paper, we study classification and regression error bounds for inhomogenous data that are independent but not necessarily identically distributed. First, we consider classification of data in the presence of non-stationary noise and establish ergodic type sufficient conditions that guarantee the achievability of the Bayes error bound, using universal rules. We then perform a similar analysis for~\(k-\)nearest neighbour regression and obtain optimal error bounds for the same. Finally, we illustrate applications of our results in the context of wireless networks.

\end{abstract}

\renewcommand{\theequation}{\thesection.\arabic{equation}}
\setcounter{equation}{0}
\section{Introduction} \label{intro}

Classification and regression are two important problems from both application and theoretical perspectives and have extensive applications in machine learning. It is well-known~\cite{dev}  that the minimum possible error is obtained by the Bayes classifier that assumes complete knowledge of the data distribution. As mentioned before, since this information is usually not available, it is of interest to compare the performance of universal rules against the Bayes classifier. When the data are independent and identically distributed (i.i.d.), it is well-known~\cite{dev}  that the~\(k-\)nearest neighbour rule is universally consistent in the sense that it achieves the Bayes bound asymptotically.  

Recently, there has been interest in studying the \emph{rate of convergence} of the error probability towards the Bayes bound and various results are known under varying assumptions on the conditional probabilities (see for example,~\cite{chau}~\cite{doring} and references therein). We also remark that inhomogenous regression has been studied from other application perspectives and in this regard, we refer to~\cite{lerch}~\cite{messner}.

In this paper, we study classification and regression of data that are not necessarily i.i.d.\ and obtain ergodic type sufficient conditions for minimizing the error probability. At the end, we discuss detailed applications of our results in the context of wireless networks.


The paper is organized as follows: In Section~\ref{sec_reg}, we obtain error bounds for regression of inhomogenous data under ergodic type conditions. Next, in Section~\ref{sec_class}, we derive bounds for the minimum classification error probability of inhomogenous data when the noise statistics satisfy an ergodic type condition. 
Finally, in Section~\ref{sec_conc} we briefly illustrate how our results are applicable in wireless networks and conclude with potential future directions.


\renewcommand{\theequation}{\thesection.\arabic{equation}}
\setcounter{equation}{0}
\section{Regression of Inhomogenous Data} \label{sec_reg}
For~\(j \geq 1\) let~\(X_{j} \in \mathbb{R}^d\) be a random vector with density~\(f_{j}\) and let~\(Y_{j} = h(X_{j})+N_j\) where~\(h\) is a deterministic but unknown function and~\(N_j\) is the noise random variable that is bounded (i.e.~\(|N_j| \leq C\) for some constant~\(C > 0\)) with zero mean and variance~\(\sigma^2_N.\) We discuss application of this setup in Section~\ref{sec_conc}, in the context of estimating transmission power levels in wireless networks.

A \emph{regressor} is a sequence of maps~\(g_n: (\mathbb{R}^d \times \mathbb{R})^{n} \times \mathbb{R}^d   \rightarrow \mathbb{R},\) that takes~\((X_j,Y_j), 1 \leq j \leq n\) and~\(X_{n+1}\) as input and outputs an estimate~\(Z_{n+1}\) of~\(Y_{n+1}.\) We say that~\(\{g_n\}\) is \emph{universal} if no~\(g_n\) depends on the distributions of~\((X_j,Y_j), j \geq 1.\) To reduce the effect of local fluctuations in the data, we use \emph{average} variance as a comparison metric and say that~\(\{g_n\}\) is \emph{consistent} if the average variance
\begin{equation}\label{err_less}
\frac{1}{n} \sum_{i=1}^{n} \mathbb{E}(Y_{i} - Z_{i})^2 \longrightarrow \sigma^2_N
\end{equation}
as~\(n \rightarrow \infty.\)

When the data is i.i.d.\ there are many well-known universally consistent regressors like classification and regression trees,~\(k-\)nearest neighbour rules (Chen and Shah (2018)) etc. Under an ergodic type condition on the data densities, we have the following result for nonstationary data.
\begin{theorem}\label{thm_reg}
If~\(h\) is uniformly continuous and
\begin{equation}\label{f_cond}
\sup_{x} \left|\frac{1}{n}\sum_{i=1}^{n}f_i(x) - f(x)\right| \longrightarrow 0
\end{equation}
as~\(n \rightarrow \infty,\) where~\(f\) is a uniformly continuous density, then there is a universally consistent regressor that achieves~(\ref{err_less}).
\end{theorem}

\emph{Proof of Theorem~\ref{thm_reg}}: Let~\({\cal S}(f) := \{x : f(x) > 0\}\) be the support of the density~\(f\) and assume for there are constants~\(0 < c_1 \leq c_2 < \infty\) such that~\(c_1 \leq f(x) \leq c_2\) for all~\(x \in {\cal S}(f).\) Also assume that~\(0 < \eta_1 \leq h(x) \leq \eta_2 < \infty\) for all~\(x.\) We relax these conditions later.

The regression problem states that given~\((X_j,Y_j), 1 \leq j \leq n\) and~\(X_{n+1},\) we are to estimate the value of~\(Y_{n+1},\)  using the~\(k-\)nearest neighbours. We begin with some prelimary computations.

Let~\(k=k(n)\) be such that~\(\frac{k}{n} \longrightarrow 0\) as~\(n \rightarrow \infty.\) For~\(x \in \mathbb{R}^d\) be such that~\(f(x) > 0\) and let~\(B_r(x)\) be the~\(d-\)dimensional ball of radius~\(r\) centred at~\(x,\) where~\(r = r(n,x) := \left(\frac{k}{\pi_d n f(x)}\right)^{1/d}\) and~\(\pi_d\) is the volume of the~\(d-\)dimensional ball of unit radius. Since~\(f(x) \geq c_1\) we get that~\(\sup_{x} r(n,x) \longrightarrow 0\) as~\(n \rightarrow \infty.\) Now if~\(N_X(x) := \sum_{i=1}^{n} \ind(X_i \in B_r(x))\) is the number of data points in~\(B_r(x),\) then~\(\mathbb{E}N_X(x) = \sum_{i=1}^{n}\int_{B_r(x)}f_i(y)dy\) and so using~(\ref{f_cond}), we get that
\begin{equation}\label{kat_tits}
\sup_{x} \left|\frac{\mathbb{E}N_X(x)}{n\pi_d r^{d}} - \int_{B_r(x)}f(y) dy\right| \longrightarrow 0
\end{equation}
as~\(n \rightarrow \infty.\) By the uniform continuity of the density~\(f\) we know that
\[\sup_{x} \left|\frac{1}{\pi_dr^{d}} \int_{B_r(x)}f(y)dy -f(x) \right| \longrightarrow 0\] as~\(n \rightarrow \infty\) and by definition~\(nf(x)\pi_dr^d = k.\) Combining these observations with~(\ref{kat_tits}), we get
\begin{equation}\label{k_conv}
\sup_{x} \left|\frac{\mathbb{E}N_X(x)}{k} -1 \right| \longrightarrow 0
\end{equation}
as~\(n \rightarrow \infty.\)

From the deviation estimate~(\ref{conc_est_f}) we also know that
\[\mathbb{P}\left(|N_X(x) - \mathbb{E}N_X(x)| \geq \epsilon \mathbb{E}N_X(x)\right) \leq \exp\left(-\frac{\epsilon^2}{4}\mathbb{E}N_X(x)\right)\]
and so from~(\ref{k_conv}), we get that
\begin{equation}\label{nx_dev}
\sup_{x} \mathbb{P}\left(|N_X(x) - k| \geq 2\epsilon k\right) \leq \exp\left(-\frac{k\epsilon^2}{8}\right)
\end{equation}
for all~\(n\) large.

Letting~\(Z_{est} := \sum_{i =1}^{n}Y_i \ind(X_i \in B_r(x))\) be the sum of the observations whose corresponding data points lie within~\(B_r(x),\)
we have that
\[\mathbb{E}Z_{est} = \sum_{i=1}^{n} \mathbb{E}Y_i\ind(X_i \in B_r(x)) = \sum_{i=1}^{n}\int_{B_r(x)}h(y)f_i(y)dy\] since the noise~\(N_i\) has zero mean and is independent of~\(X_i,\) for each~\(i.\) Again using~(\ref{f_cond}) and the uniform continuity of~\(h(x),\) we then get that
\[\sup_{x} \left|\frac{1}{n\pi_dr^{d}} \sum_{i=1}^{n}\int_{B_r(x)}h(y)f_i(y)dy - h(x)f(x)\right| \longrightarrow 0\]  as~\(n \rightarrow \infty\) and so an analogous argument as in~(\ref{k_conv}) gives us
\begin{equation}\label{k_y_conv}
\sup_{x} \left| \frac{\mathbb{E}Z_{est}}{k} -h(x) \right| \longrightarrow 0
\end{equation}
as~\(n \rightarrow \infty.\) The function~\(h(.)\) and the noise~\(N_i\) are both bounded and therefore each~\(|Y_i| \leq B\)  for some constant~\(B>  0.\) This allows us to use the Azuma-Hoeffding inequality (Alon and Spencer (2008)) to obtain that
\begin{eqnarray} \label{ny_est_dev}
\sup_{x} \mathbb{P}\left(|Z_{est} - kh(x)| \geq 2\epsilon k h(x)\right) &\leq& \sup_{x} \exp\left(-C_1\epsilon^2 k h(x) \right) \nonumber\\
&\leq& \exp\left(-C_2\epsilon^2 k \right),
\end{eqnarray}
for some constants~\(C_i = C_i(\eta_1,\eta_2) > 0,\) where the final relation in~(\ref{ny_est_dev}) is true since~\(h(x) \geq \eta_1 > 0.\)

Defining
\[E_{tot}(x) := \left\{|N_X(x) - k| \leq 2\epsilon k\right\} \bigcap \left\{|Z_{est} - kh(x)| \leq 2\epsilon k h(x)\right\},\]
we get from~(\ref{nx_dev}),~(\ref{ny_est_dev}) and the union bound that
\begin{equation}\label{e_tot_est}
\inf_{x} \mathbb{P}(E_{tot}(x)) \geq 1-\exp\left(-\frac{k\epsilon^2}{8}\right) -\exp\left(-C_2\epsilon^2 k\right) \geq 1-e^{-C_3 \epsilon^2 k}
\end{equation}
for some constant~\(C_3 > 0.\) We assume henceforth that~\(E_{tot}(x)\) occurs and let~\({\cal N}_k(x)\) be the~\(k\) closest data points to~\(x\) in the set~\(\{X_i\}_{1 \leq i \leq n}\)  and define
\begin{equation}\label{knn_def}
Y_{kNN} := \frac{1}{k}\sum_{i=1}^{n} Y_i \ind(X_i \in {\cal N}_k(x))
\end{equation}
to be the corresponding sum of the observations. Because~\(E_{tot}(x)\) occurs, the number of data points within~\(B_r(x)\) lies between~\(k-k\epsilon\) and~\(k+k\epsilon\) and so using~\(|Y_i| \leq B\) for each~\(i,\) we see that
\[\frac{1}{k}Z_{est} - B\epsilon \leq Y_{kNN} \leq \frac{1}{k} Z_{est} + B\epsilon\] and therefore that
\[h(x) - 3B\epsilon \leq Y_{kNN} \leq h(x) + 3B\epsilon.\]

Consequently, we get from~(\ref{ny_est_dev}) that
\begin{equation}\label{knn_bound_ax}
\sup_{x} \mathbb{P}\left(|Y_{kNN}-h(x)| \geq 3B\epsilon\right) \leq e^{-C_3 \epsilon^2 k}
\end{equation}
and therefore that
\begin{equation}\label{knn_bound}
\sup_{x} \mathbb{E}\left|Y_{kNN} - h(x)\right|^2 \leq 9B^2\epsilon^2 + e^{-C_3 \epsilon^2 k}
\end{equation}
Setting~\(\epsilon= \frac{(\log{k})}{\sqrt{k}}\) the right side of~(\ref{knn_bound}) is at most
\[\frac{9B^2(\log{k})^2}{k}  + \exp\left(-C_3(\log{k})^2 \right) \leq \frac{C_4(\log{k})^2}{k} =: b_n(k) = b_n\] for all~\(n\) large and some constant~\(C_4 > 0.\) Therefore using the mutual independence of~\(Y_{kNN}\) and~\(N_{n+1}\) and the fact that~\(N_{n+1}\) has zero mean, we get that
\[\mathbb{E}(Y_{kNN}-h(x)-N_{n+1})^2 = \mathbb{E}(Y_{kNN}-h(x))^2 + \mathbb{E}N_{n+1}^2 \leq b_n + \sigma_N^2.\] Finally, using Fubini's theorem, we see that the average variance as defined in~(\ref{err_less}) is at most~\(\frac{1}{n}\sum_{i=1}^{n} b_i + \sigma_N^2 \longrightarrow \sigma_N^2\)
as~\(n \rightarrow \infty,\) since~\(b_n \longrightarrow 0\) as~\(n \rightarrow \infty.\)

Next we relax the condition on the lower and upper bounds for the density~\(f.\) For~\(\zeta > 0\) small, we define the open set
\[A_{good} = \left\{\zeta < f < \zeta^{-1}\right\} \bigcap 
\left\{\zeta < h < \zeta^{-1}\right\} \] and perform an analogous analysis as above for~\(x \in A_{good}.\) Using~(\ref{knn_bound}) and arguing as above, we then get that
\[\mathbb{E}|Y_{kNN}-Y_{n+1}|^2 \leq \sigma_N^2 + b_n + \int_{A^c_{good}} f_{n+1}(x)dx\] and so the average variance is at most
\[\sigma_N^2 + \frac{1}{n}\sum_{i=1}^{n}b_i + \int_{A^c_{good}} \frac{1}{n}\sum_{i=1}^{n}f_i(x) dx.\] Again using~(\ref{f_cond}) and the fact that~\(b_n \longrightarrow 0,\) we get that the average variance is at most~\[\sigma_N^2 + \int_{A^c_{good}} f(x)dx \longrightarrow \sigma_N^2\] as~\(\zeta \rightarrow 0.\)~\(\qed\)

\renewcommand{\theequation}{\thesection.\arabic{equation}}
\setcounter{equation}{0}
\section{Classification of Inhomogenous Data} \label{sec_class}
Let~\(\mu\) be a probability measure in~\(\mathbb{R}^d\) and let~\(X\) be a random variable with measure~\(\mu.\)  Let~\(X_i, 1 \leq i \leq n\) be independent and identically distributed (i.i.d.) with the same distribution as~\(X.\) For~\(1 \leq i \leq n\) let~\(Y_i =h_i(X_i) \in \{0,1\}\) be a binary random variable depending only on~\(X_i\) and independent of all~\((X_j,Y_j), j \neq i.\) We discussion application of this setup in Section~\ref{sec_conc} in the context of primary user detection in cognitive radio networks.

The random variables~\(\{Y_i\}\) need \emph{not} be identically distributed and we refer to~\((X_i,Y_i), 1 \leq i \leq n\) henceforth as the \emph{data}. From the Fubini's theorem, we have that
\begin{equation}\label{cond_prob_def}
\mathbb{P}(Y_i=1, X_i \in A) = \int_{A} h_i(x) \mu(dx)
\end{equation}
for all measurable sets~\(A.\)


A classification rule is a sequence of deterministic measurable functions~\(g_n, n \geq 1\) such that~\(g_n : \mathbb{R}^{2nd+1} \rightarrow \{0,1\}\) takes~\((X_i,Y_i), 1 \leq i \leq n\) and~\(X_{n+1}\) as input and outputs its prediction of~\(Y_{n+1}.\) We say that the rule~\(\{g_n\}\) is \emph{universal} if no~\(g_n\) depends either on~\(\mu\) or any of the~\(h_i.\) For e.g., the~\(k-\)nearest neighbours rule and the histogram rule~\cite{dev}  are examples of universal classifiers. Formally, the~\(k-\)NN rule is described as follows. For each point~\(x,\) let~\(N_1\) be the number of~\(1's\) amongst the~\(k\) nearest neighbours of~\(x,\) in terms of Euclidean distance. We then define the~\(k-\)NN rule as
\begin{equation}\label{knn_rule}
w_n(x) := \left\{
\begin{array}{ll}
0 & \text{ if } N_1 \leq \frac{k}{2}\\
1 & \text{ otherwise}.
\end{array}
\right.
\end{equation}

For any classification rule~\(\{g_n\},\) we define~\[\varepsilon_n := \mathbb{P}\left(Y_{n+1} \neq g_n(X_{n+1})\right)\] to be the error probability at the~\((n+1)^{th}\) data point, where~\(g_n(X_{n+1}) = g_n\left(X_{n+1};X_1,Y_1,\ldots,X_{n},Y_{n}\right)\) possibly depends on all or some of the previous data points. In general, it is of interest to minimize~\(\varepsilon_n\) and if~\(\{Y_i\}\) are i.i.d., i.e.~\(Y_i = h(X_i)\) for some deterministic function~\(h,\) then it is well-known (Theorem~\(2.1,\) pp.~\(10,\)~\cite{dev}) that the desired minimization is achieved using the Baye's classification rule: i.e.~\(g_n = \ind\left(h > \frac{1}{2}\right)\) where~\(\ind(.)\) refers to the indicator function. In this case
\begin{equation}\label{err_prob_ax}
\varepsilon_n = \int \min(h(x),1-h(x)) \mu(dx) =: L^*
\end{equation}
and the minimum error remains invariant, i.e. does not change with the index~\(n.\)

If~\(\{Y_i\}\) are not identically distributed, then the minimum error might fluctuate, i.e. vary with~\(n.\) To counter the effect described above, in this paper, we define and study the \emph{average} error probability
\begin{equation}\label{ave_err_prob}
T_n := \frac{1}{n}\sum_{i=1}^{n} \varepsilon_i.
\end{equation}
Also, because it may not be feasible to estimate~\(\eta_i\) for each~\(i,\) we define our error probabilities of interest as
\begin{equation}\label{err_prob}
E^* := \inf_{\{g_n\}}\liminf_{n} T_n \text{ and } U^* = \inf_{\{u_n\}}\limsup_{n} T_n
\end{equation}
where the infimum for~\(E^*\) is over all possible classification rules and the infimum for~\(U^*\) is over all possible \emph{universal} classification rules.  We say that a universal classification rule~\(\{u_n\}\) is \emph{universally consistent}~\cite{dev} if it achieves~\(U^*,\) i.e. the error probability~\(\limsup_{n}T_n = U^*.\)

Our main result below obtains the minimum possible average error probability under an ergodic type condition on the noise statistics~\(\{\eta_i\}.\)
\begin{theorem} \label{thm_ax}
If
\begin{equation}\label{conv_cond_ax22}
\sup_{x} \frac{1}{n}\sum_{i=1}^{n} |h_i(x) - h(x)| \longrightarrow 0
\end{equation}
then~\(E^* =U^*=L^*,\) where~\(L^*\) is the Bayes classification error as defined in~(\ref{err_prob_ax}). If in addition
\begin{equation}\label{n_tits}
\mathbb{P}\left(h(X) = \frac{1}{2}\right) = 0,
\end{equation}
then there is a universally consistent rule~\(\{u_n\}\) that achieves~\(L^*.\)
\end{theorem}
Condition~\ref{conv_cond_ax22}) is an ergodic type condition for the existence of universal classifiers performing arbitrarily close to optimality as possible. The stronger condition~(\ref{n_tits}) states that as long as the output~\(Y\) is never ``truly" random, there are universal classifiers that in fact achieve the Bayes bound~\(L^*.\)


In our proof of Theorem~\ref{thm_ax} below, we use a majority rule to estimate the error probabilities and obtain the bound~\(U^* \leq L^*.\) This is described in the following auxiliary result of independent interest.
\begin{lemma}\label{aux_lem} If
\begin{equation}\label{conv_cond_ax}
\sup_{x} \left| \frac{1}{n}\sum_{i=1}^{n} h_i(x) - h(x) \right| \longrightarrow 0,
\end{equation}
then for every~\(\zeta > 0\) there is a sequence of universal classifiers~\(\{g_n\}\) whose error probability satisfies~\(\limsup_nT_n \leq L^* + b(\zeta)\) where~\(b(\zeta) \longrightarrow 0\) as~\(\zeta \rightarrow 0.\)
\end{lemma}

\emph{Proof of Lemma~\ref{aux_lem}}: We use a classifier based on the following majority rule: Let~\(r = r_n > 0\) be such that
\begin{equation}\label{rn_conv}
r_n \rightarrow 0 \text{ and } \frac{nr_n^d}{\log{n}} \longrightarrow \infty,
\end{equation}
as~\(n \rightarrow \infty\)  and for~\(x \in \mathbb{R}^d,\) let~\(S_r(x)\) be the Euclidean ball of radius~\(r\) centred at~\(x.\) Let~\(N_{tot}\) be the total number of data points of~\(\{X_i\}_{1 \leq i \leq n},\) present within~\(S_r(x)\) and for~\(k \in \{0,1\}\) let~\(N_{k}\) be the number of~\(k's\) amongst the data points~\(\{Y_i\}_{1 \leq i \leq n},\) within~\(S_r(x).\)  For~\(0 < \zeta < \frac{1}{4},\) we define our classification rule as
\begin{equation}\label{gn_def_ax}
g_n(x) := \left\{
\begin{array}{ll}
1, & \text{ if } N_1-N_0 > \zeta N_{tot}\\
Z_n, & \text{ if } |N_1-N_0| \leq 2\zeta^2 N_{tot}\\
0, & \text{ otherwise },
\end{array}
\right.
\end{equation}
where~\(\{Z_j\}\) are i.i.d. random binary random variables also independent of the data~\(\{(X_i,Y_i)\},\) satisfying~\[\mathbb{P}(Z_j = 1 ) = \frac{1}{2} = \mathbb{P}(Z_j = 0).\] In words, as long as the number of~\(k's\) is reasonably larger than the number of~\((1-k)'s\) our classifier outputs~\(k \in \{0,1\}\) as the predicted value. Otherwise, our classifier plays it safe and outputs the result a random coin toss in the form of~\(Z_n.\)

Throughout, we use the following standard deviation estimate which we state separately for convenience:  Let~\(\{Z_j\}_{1 \leq j \leq r}\) be independent Bernoulli random variables with~\(\mathbb{P}(Z_j = 1) = 1-\mathbb{P}(Z_j = 0) > 0.\) If~\(U_r := \sum_{j=1}^{r} Z_j, \theta_r := \mathbb{E}U_r\) and~\(0 < \gamma \leq \frac{1}{2},\) then
\begin{equation}\label{conc_est_f}
\mathbb{P}\left(\left|U_r - \theta_r\right| \geq \theta_r \gamma \right) \leq 2\exp\left(-\frac{\gamma^2}{4}\theta_r\right)
\end{equation}
for all \(r \geq 1.\) For a proof of~(\ref{conc_est_f}), we refer to Corollary A.1.14, pp. 312 of Alon and Spencer (2008).

We prove the desired result in three steps.   In the first step, we prove certain convergence results used in the final evaluation of the error probability and in the second step, we derive analytically tractable expressions for the error probability. In our third step, we combine the results of Steps~\(1\) and~\(2\) to obtain the desired bound on~\(U^*.\)

\emph{\underline{Step 1}}: For~\(\mu-\)a.e.~\(x,\) we show that
\begin{equation}\label{conv_prop_main}
\frac{N_1}{N_{tot}} \longrightarrow \eta(x)\text{ a.s.\ }
\end{equation}
as~\(n \rightarrow \infty.\)   To prove~(\ref{conv_prop_main}), we use the following estimates that hold~\(\mu-\)a.e.~\(x:\)
\begin{equation}\label{meas_prop}
\lim_{r \rightarrow 0} \frac{1}{\mu(S_r(x))} \int_{S_r(x)} h d\mu  = h(x) \text{ and } \lim_{r \rightarrow 0} \frac{r^d}{\mu(S_r(x))} < \infty.
\end{equation}
The first result in~(\ref{meas_prop}) is the generalization of Lebesgue differentiation theorem for Radon measures (see Theorem~\(11.1,\) pp. 192 in DiBenedetto (2002)) and the second relation in~(\ref{meas_prop}) is a consequence of the almost sure finiteness of the derivatives of Radon measures (set~\(\nu\) to be the Lebesgue measure in Proposition~\(9.1,\) pp. 186 and Equation~\(9.1,\) pp. 187 of DiBenedetto (2002)).

Using~(\ref{meas_prop}), we obtain~(\ref{conv_prop_main}) as follows. By definition~\[\mathbb{E}N_{tot} = n \mu(S_r(x)) =: n\mu_r\] and so from the deviation estimate~(\ref{conc_est_f}), we get
\begin{eqnarray}
\mathbb{P}((1-\epsilon)n\mu_r \leq N_{tot} \leq (1+\epsilon)n\mu_r) &\geq& 1-\exp\left(-C_1n\mu_r\right) \nonumber\\
&\geq& 1-\exp\left(-C_2nr^d\right) \label{thmud3}
\end{eqnarray}
for some constants~\(C_1,C_2 > 0,\) by the second relation in~(\ref{meas_prop}). From the second relation in~(\ref{rn_conv}), we get that~\(C_2nr^d \geq 2\log{n}\) for all~\(n\) large and so the Borel-Cantelli Lemma implies that
\begin{equation}\label{rel_tot}
\frac{N_{tot}}{\mathbb{E}N_{tot}} = \frac{N_{tot}}{n\mu_r} \longrightarrow 1 \text{ a.s.\  }
\end{equation}
as~\(n \rightarrow \infty.\)

Next we have that~\(\mathbb{E}N_1 = \int_{S_r(x)} \sum_{i=1}^{n} h_i d\mu\) and so using the uniform convergence condition~(\ref{conv_cond_ax}),
we have for~\(\epsilon > 0\) that
\begin{eqnarray}
\left|\frac{1}{n\mu_r}\mathbb{E}N_1 - \frac{1}{\mu_r}\int_{S_r(x)}h d\mu\right| &=& \left|\frac{1}{\mu_r}\int_{S_r(x)} \left(\frac{1}{n} \sum_{i=1}^{n} h_i -h \right) d\mu \right| \nonumber\\
&\leq& \frac{1}{\mu_r}\int_{S_r(x)} \left|\frac{1}{n}\sum_{i=1}^{n} h_i - h\right| d\mu \nonumber\\
&\leq& \epsilon
\end{eqnarray}
for all~\(n\) large. Combining this with the first relation in~(\ref{meas_prop}), we then get that
\begin{equation}\label{temp_ax}
\frac{\mathbb{E}N_1}{n\mu_r} \longrightarrow h(x)
\end{equation}
as~\(n \rightarrow \infty.\)

If~\(h(x) > 0\) then
\begin{equation}\label{dilpa}
\mathbb{E}N_1 \geq D_1n\mu_r \geq D_2 nr^d
\end{equation}
for all~\(n\) large and some constants~\(D_1,D_2 > 0,\) where the second inequality in~(\ref{dilpa}) is true by the second condition in~(\ref{meas_prop}). Consequently, the deviation estimate~(\ref{conc_est_f}) gives us that
\begin{equation}\label{dilpa2}
\mathbb{P}\left(\mathbb{E}N_1(1-\epsilon) \leq N_1 \leq \mathbb{E}N_1(1+\epsilon)\right) \geq 1-\exp\left(-D_3nr^d\right) \geq 1-\frac{1}{n^2}
\end{equation}
for some constant~\(D_3 > 0,\) where the final estimate in~(\ref{dilpa2}) is true by the second condition in~(\ref{rn_conv}). Since~\(\epsilon > 0\) is arbitrary, the Borel-Cantelli Lemma then implies that~\(\frac{N_1}{\mathbb{E}N_1} \longrightarrow 1\) a.s.\ as~\(n \rightarrow \infty\) and combining this with~(\ref{temp_ax}) and~(\ref{rel_tot}), we get~(\ref{conv_prop_main}).

If~\(h(x) = 0,\) then we use Chernoff bound to obtain the almost sure convergence. Indeed, letting~\(p_i := \mathbb{P}(Y_i=1,X_i \in S_r(x)),\) we have that
\[\mathbb{E}e^{N_1} = \prod_{i=1}^{n} (1+(e-1)p_i)\leq \exp\left((e-1)\sum_{i}p_i\right) = \exp\left((e-1) \mathbb{E}N_1\right).\]
For any~\(\epsilon > 0,\) we use~(\ref{temp_ax}) and the second relation in~(\ref{meas_prop}) to get that\\\(\mathbb{E}N_1 \leq \epsilon nr^d\)  for all~\(n\) large. Consequently, the Chernoff bound gives us that
\[\mathbb{P}(N_1 \geq 2\epsilon nr^d) \leq e^{-2\epsilon nr^d} \mathbb{E}e^{N_1} \leq e^{-\epsilon nr^d}\]
and so again using the second conditions in~(\ref{meas_prop}) and~(\ref{rn_conv}), we get that
\[\mathbb{P}(N_1 \geq C\epsilon n\mu_r) \leq \frac{1}{n^2}\] for some constant~\(C >0.\)
As before, the Borel-Cantelli Lemma and the fact that~\(\epsilon > 0\) is arbitrary implies that~\(\frac{N_1}{n\mu_r}  = \frac{N_1}{\mathbb{E}N_{tot}} \longrightarrow 0\) a.s. and so from~(\ref{rel_tot}), we get~(\ref{conv_prop_main}).

\emph{\underline{Step 2}}: In this step, we obtain an expression for~\(T_n\) in terms of the classifier error probability. We first rewrite~\(\varepsilon_n\) in terms of the classifier outputs~\(g_n(.)\) as
\begin{equation}\label{var_exp2}
\varepsilon_n = \mathbb{P}(Y_{n+1} \neq g_n(X_{n+1})) = \mathbb{P}(Y_{n+1} = 1,g_n=0) + \mathbb{P}(Y_{n+1} =0, g_n = 1).
\end{equation}
By Fubini's theorem we have
\[\mathbb{P}(Y_{n+1}=1,g_n=0) = \int h_{n+1}(x) \mathbb{P}(g_n(x)= 0) \mu(dx)\] and
\[\mathbb{P}(Y_{n+1}=0,g_n=1) = \int (1-h_{n+1}(x)) \mathbb{P}(g_n(x)= 1) \mu(dx).\]
Plugging these into~(\ref{var_exp2}) we get that
\begin{equation}\label{var_ep_exp}
\varepsilon_n = \int \theta_n(x) \mu(dx),
\end{equation}
where
\begin{equation}\label{theta_n_def}
\theta_n(x) : = h_{n+1}(x) \mathbb{P}(g_n(x)= 0) + (1-h_{n+1}(x)) \mathbb{P}(g_n(x)= 1)
\end{equation}
Consequently,
\begin{equation}\label{tn_expo}
T_n = \int\frac{1}{n}\sum_{i=1}^{n} \theta_i(x) \mu(dx).
\end{equation}

\emph{\underline{Step 3}}: We now combine the results of Steps~\(1\) and~\(2\) to complete the proof of the Lemma. From~(\ref{conv_prop_main}), we see that~\(\frac{N_1-N_0}{N_{tot}} \longrightarrow 2h(x)-1\) a.s.\ as~\(n \rightarrow \infty.\) Therefore if~\(h(x)   > \frac{1+2\zeta}{2},\) then \[N_1-N_0 >(2h(x)-1-\zeta^2) N_{tot} > (2\zeta-\zeta^2) N_{tot} > \zeta N_{tot}\] for all~\(n\) large a.s.\ since~\(0 < \zeta < \frac{1}{4}.\) Therefore we get that
\begin{equation}\label{thilp}
\mathbb{P}(g_n(x) =1) \longrightarrow 1.
\end{equation}
Similarly if~\(h(x) < \frac{1-2\zeta}{2}\) then~\[N_0-N_1 >(1-2h(x)-\zeta^2) N_{tot} > 2\zeta-\zeta^2 N_{tot} > 2\zeta^2 N_{tot}\] for all~\(n\) large a.s.\ and so
\begin{equation}\label{thilp22}
\mathbb{P}(g_n(x) =0) \longrightarrow 1
\end{equation}
as~\(n \rightarrow \infty.\) Finally, if~\(|2h(x)-1| \leq \zeta^2\) then again using~(\ref{conv_prop_main}) we get that\\\(|N_1-N_0| \leq 2\zeta^2 N_{tot}\)  for all~\(n\) large a.s. Thus~\(\mathbb{P}(g_n = Z_n \text{ eventually}) \longrightarrow 1\) as~\(n \rightarrow \infty\) and so~\(\mathbb{P}(g_n = Z_n) \longrightarrow 1\) as~\(n \rightarrow \infty.\) Consequently
\[\mathbb{P}(g_n=0) = \mathbb{P}(Z_n=0) + \mathbb{P}(g_n=0,g_n \neq Z_n) - \mathbb{P}(Z_n=0,g_n \neq Z_n) \longrightarrow \frac{1}{2}\]
as~\(n\rightarrow \infty\) and hence
\begin{equation}\label{thilp3}
\frac{1}{n}\sum_{i=1}^{n}\theta_i(x) \longrightarrow \frac{1}{2}
\end{equation}
as~\(n \rightarrow \infty.\)

Now if~\(\{a_n\}\) and~\(\{b_n\}\) are bounded sequences satisfying
\[a_n \rightarrow a \text{ and } \frac{1}{n}\sum_{i=1}^{n}b_i \longrightarrow b,\] then
\begin{equation}\label{conv_ab}
\frac{1}{n} \sum_{i=1}^{n} a_ib_i \longrightarrow ab.
\end{equation}
Therefore plugging~(\ref{thilp}),~(\ref{thilp22}) and~(\ref{thilp3}) into~(\ref{theta_n_def}) and using the convergence condition~(\ref{conv_cond_ax}), we get for~\(\mu-\)a.e.~\(x\) that
\[\lim_n \frac{1}{n}\sum_{i=1}^{n}\theta_i(x)
=
\left\{
\begin{array}{ll}
1-h(x), & \text{ if } h(x) > \frac{1}{2} + \zeta\\
h(x), & \text{ if } h(x) < \frac{1}{2} - \zeta\\
\frac{1}{2}, & \text{ if } \frac{1}{2} - \zeta^2 < h(x)  < \frac{1}{2} + \zeta^2.
\end{array}
\right.
\]
From~(\ref{tn_expo}) and Fatou's Lemma, we then obtain~\(\limsup_n T_n \leq L^* + b(\zeta),\)
where
\[b(\zeta) := \mathbb{P}\left(\eta(X) \in \left[\frac{1}{2}-\zeta,\frac{1}{2}\right) \cup \left(\frac{1}{2},\frac{1}{2} + \zeta\right] \right) \longrightarrow 0\]
as~\(\zeta \rightarrow 0.\)~\(\qed\)

\emph{Proof of Theorem~\ref{thm_ax}}: We first show that~\(\liminf_{n}T_n \geq L^*.\) From~(\ref{err_prob_ax}), we immediately see that
\[\liminf_{n} T_n \geq M^* := \liminf_{n} \frac{1}{n} \sum_{i=1}^{n} \mathbb{E}\min(h_i,1-h_i).\] Using~\(\min(a,b) = \frac{a+b-|a-b|}{2} \) we get that
\begin{eqnarray}
M^* &=& \liminf_{n}\left(\frac{1}{2} - \frac{1}{2n}\sum_{i=1}^{n} \mathbb{E}|2h_i-1|\right) \nonumber\\
&=& \frac{1}{2} - \limsup_{n} \frac{1}{n}\sum_{i=1}^{n} \mathbb{E}\left|h_i-\frac{1}{2}\right|. \label{m_bd}
\end{eqnarray}
Similarly writing~\(L^* = \frac{1}{2} - \mathbb{E}\left|h-\frac{1}{2}\right|\) it suffices to show that
\[\frac{1}{n}\sum_{i=1}^{n} \left|h_i-\frac{1}{2}\right| \longrightarrow \left|h-\frac{1}{2}\right|\] as~\(n \rightarrow \infty.\) Indeed, using~\(\left||a|-|b|\right| \leq |a-b|\) we get that
\[\left| \left|h_i-\frac{1}{2}\right|  - \left|h-\frac{1}{2}\right| \right| \leq |h_i-h|\] and so using the convergence condition~(\ref{conv_cond_ax22}) and the bounded convergence theorem, we see that~\(M^* = L^*.\)

If~(\ref{n_tits}) holds, then we define our universal classification rule as
\begin{equation}\label{gn_def_ax2}
g_n(x) := \left\{
\begin{array}{ll}
0, & \text{ if } N_1 \leq \frac{N_{tot}}{2}\\
1, & \text{ otherwise}.
\end{array}
\right.
\end{equation}
From~(\ref{conv_prop_main}) in the proof of Lemma~\ref{aux_lem}, we see that if~\(\eta(x)  \neq \frac{1}{2},\) then
\begin{equation}\label{thilp_ax}
\mathbb{P}(g_n(x) =0) \longrightarrow \ind\left(h(x) < \frac{1}{2}\right) \text{ and } \mathbb{P}(g_n(x) = 1) \longrightarrow \ind\left(h(x) > \frac{1}{2}\right).
\end{equation}
Therefore from the convergence condition~(\ref{conv_cond_ax}),~(\ref{thilp_ax}),~(\ref{theta_n_def}) and~(\ref{conv_ab}), we see that if~\(h(x) \neq \frac{1}{2},\) then
\begin{equation}\label{thmud2}
\frac{1}{n}\sum_{i=1}^{n}\theta_i(x) \longrightarrow h(x) \ind\left(h(x) < \frac{1}{2}\right)  + (1-h(x)) \ind\left(h(x) > \frac{1}{2}\right).
\end{equation}
Plugging~(\ref{thmud2}) into~(\ref{tn_expo}) and using Fatou's Lemma and~(\ref{n_tits}), we get the desired bound\\\(\limsup_n T_n \leq L^*.\)~\(\qed\)

\renewcommand{\theequation}{\thesection.\arabic{equation}}
\setcounter{equation}{0}
\section{Applications and Future Directions} \label{sec_conc}
In this section, we briefly describe applications of our results in the context of wireless networks. 

\subsection*{Regression Application} 
Consider a typical wireless network scenario, where mobile nodes are distributed randomly in a geographical region and each node is transmitting data to the base station at a certain power level that depends on the location of the node. Farther the node from the base station, larger is the transmitting power required to maintain a minimum quality of service (QoS). 

In the above context, the power level~\(Y_j\) of the~\(j^{th}\) node is a function~\(h(X_j)\) of the node location~\(X_j\) and in practical situations where measurement of power level has small error, we can in fact assume that there is no noise, 
i.e.~\(N_j =0.\) 

Now if there were no shadowing or fading, then we can set~\(h(x) = \frac{1}{x^{\delta}}\) for some~\(\delta > 2\) (see~\cite{goldsmith}, Chapter~\(2\) for more details). However, in practice channel fading and shadowing affect the received power level significantly (Chapter~\(3,\)~\cite{goldsmith}) and so usually the received power level is a deterministic function~\(h\) that solely depends on the environment.

Suppose now a new node enters the network and would like to interact and transmit data via the base station. Because, it has no \emph{a priori} knowledge of the environment, it could gather information about the transmitting power levels of its~\(k-\)nearest neighbours for some suitably chosen~\(k\) and estimate its own required transmission power using regression. In other words, the new node ``discovers" its surrounding environment using regression.

\subsection*{Classification Application}
We demonstrate how our classification error bounds have application in cognitive radio (CR) networks consisting of unlicensed users that are allowed to use licensed spectrum with the understanding that they vacate upon arrival of the licensed user~\cite{ganesan}.

Unlicensed users ``far" from the licensed user could continue to use the spectrum since they would not cause any interference. However, unlicensed users close to the licensed user must vacate. Therefore in this context, we let~\(h_i(.)\) denote the indicator function that is one if~\(i^{th}\) user is a disturbance to the licensed user and is zero otherwise. Again~\(h_i(.)\) depends on the geography and also the behaviour of the~\(i^{th}\) unlicensed user (like if it is transmitting at high power levels etc).

As before, a new unlicensed user could use neighbourhood information to determine whether or not it would be a disturbance to the licensed user and proceed to either transmit or not, based on the decision outcome.

\subsection*{Future Directions}
In this paper, we have considered regression and classification for inhomogenous data with applications to wireless networks. We have also discussed the applications above. It would be interesting to relax the assumptions in our main Theorems and see how performance degrades as a result. Moreover, it would be even interesting to obtain partial or full converses that would essentially demonstrate the optimality of our results.

\subsection*{\em Acknowledgement}
I thank Professors Rahul Roy, Thomas Mountford, Federico Camia, Alberto Gandolfi and C. R. Subramanian for crucial comments and also thank IMSc and IISER Bhopal for my fellowships.

\end{document}